\documentclass[runningheads]{llncs}
\usepackage[T1]{fontenc}
\usepackage{graphicx}
\usepackage{xcolor}
\usepackage{array, makecell}
\usepackage[utf8]{inputenc}
\usepackage{enumitem}
\usepackage{adjustbox}
\usepackage{longtable}
\usepackage{multicol}
\usepackage{supertabular}
\usepackage{multirow}
\usepackage{booktabs}
\usepackage{array, makecell}
\usepackage{amsmath}
\usepackage{calrsfs}
\usepackage{amsfonts}
\DeclareMathAlphabet{\pazocal}{OMS}{zplm}{m}{n}

\newcommand{\Lb}{\pazocal{L}}

\begin{document}
%
\title{Parameter-Efficient Sparse Retrievers and Rerankers using Adapters}

%
%
\author{Vaishali Pal\inst{1,2}\orcidID{0000-0002-1493-3659} \and
Carlos Lassance\inst{2}\orcidID{0000-0002-7754-6656} \and
Hervé Déjean\inst{2}\orcidID{0000-0002-9837-5358} \and
Stéphane Clinchant\inst{2} \orcidID{0000-0003-2367-8837}}
\authorrunning{V. Pal et al.}
%
\institute{IRLab, University of Amsterdam, Amsterdam, Netherlands  \and
 Naver Labs Europe, Meylan, France}
%
\maketitle              
\begin{abstract}
Parameter-Efficient transfer learning with Adapters have been studied in Natural Language Processing (NLP) as an alternative to full fine-tuning. Adapters are memory-efficient and scale well with downstream tasks by training small bottle-neck layers added between transformer layers while keeping the large pretrained language model (PLMs) frozen. In spite of showing promising results in NLP, these methods are under-explored in Information Retrieval. While previous studies have only experimented with dense retriever or in a cross lingual retrieval scenario, in this paper we aim to complete the picture on the use of adapters in IR. First, we study adapters for SPLADE, 
a sparse retriever, for which adapters not only retain the efficiency and effectiveness otherwise achieved by finetuning, but are memory-efficient and orders of magnitude lighter to train. 
We observe that Adapters-SPLADE not only optimizes just 2\% of training parameters, but outperforms fully fine-tuned counterpart and existing parameter-efficient dense IR models on IR benchmark datasets.
Secondly, we address domain adaptation of neural retrieval thanks to adapters 
on cross-domain BEIR datasets and TripClick.
Finally, we also consider knowledge sharing between rerankers and first stage rankers. Overall, our study complete the examination of adapters for neural IR.\footnote{The code can be found at:https://github.com/naver/splade/tree/adapter-splade}


\end{abstract}
\section{Introduction}
Information Retrieval (IR) systems often aim to return a ranked list of documents ordered with respect to their relevance to a user query. 
In modern web search engines, there is, in fact, not a single retrieval model but several ones specialized in diverse information needs such as different search verticals.
To add to this complexity, multi-stage retrieval considers effectiveness-efficiency trade-off where first stage retrievers are essential for fast retrieval of potentially relevant candidate documents from a large corpus. Further down the pipeline, rerankers are added focusing on effectiveness. 

With the advent of large Pretrained Language Models (PLM), recent neural retrieval models have millions of parameters. Training, updating and adapting such models implies significant computing and storage cost calling for efficient methods. Moreover, generalizability across out-of-domain datasets is critical and even when effectively adapted to new domains, full finetuning often comes at the expense of large storage and catastrophic forgetting. Fortunately, such research questions have already been studied in the NLP literature \cite{Beck2022AdapterHubPS,ben-zaken-etal-2022-bitfit,conf/icml/HoulsbyGJMLGAG19,hu2021lora} with parameter-efficient tuning. In spite of very recent work exploring parameter-efficient techniques for neural retrieval, the use of adapters in IR has been overlooked. Previous work on dense retriever had mixed results~\cite{10.1145/3485447.3511978} and successful adaptation was achieved for cross lingual retrieval~\cite{DBLP:journals/corr/abs-2204-02292}. Our study aims to complete the examination of adapters for neural IR and investigates it with neural sparse retrievers. We study ablation of adapter layers to analyze whether all layers contribute equally. We examine how adapter-tuned neural sparse retriever SPLADE~\cite{10.1145/3477495.3531857} fares on benchmark IR datasets MS MARCO~\cite{nguyen2016ms}, TREC DL 2019 and 2020~\cite{craswell2021trec} and out-of-domain BEIR datasets~\cite{thakur2021beir}. We explore whether generalizability of SPLADE can be further improved with adapter-tuning on BEIR and out-of-domain dataset such as TripClick~\cite{rekabsaz2021tripclick}. In addition, we examine knowledge transfer between first stage retrievers and rerankers with full fine-tuning and adapter-tuning. To the best of our knowledge, this is the first work which studies adapters on sparse retrievers, focuses on sparse models' generalizability and explores knowledge transfer between retrievers in different stages of the retrieval pipeline. In summary, we address the following  research questions:
\begin{enumerate}
    \item RQ1: What is the efficiency-accuracy trade-off of parameter-efficient fine-tuning with adapters on the sparse retriever model SPLADE?
    \item RQ2: How does each adapter layer ablation affect  retrieval effectiveness?
    \item RQ3: Are adapters effective for adapting neural sparse neural retrieval in a new domain?
    \item RQ4: Could adapters be used to share knowledge between rerankers and first stage rankers?
\end{enumerate}

\section{Background and Related Work}
Parameter efficient transfer learning techniques aim to adapt large pretrained models to downstream tasks using a fraction of training parameters, achieving comparable effectiveness to full fine-tuning. Such methods \cite{conf/icml/HoulsbyGJMLGAG19,li-liang-2021-prefix,hu2021lora,Pfeiffer2020MADXAA,ruckle-etal-2021-adapterdrop} are memory efficient and scale well to numerous downstream tasks due to the massive reduction in task specific trainable parameters. This makes them an attractive solution for efficient storage and deployment compared to fully fine-tuned instances.
Such methods have been successfully applied to language translation \cite{Pfeiffer2020MADXAA}, natural language generation \cite{lin-etal-2020-exploring}, Tabular Question Answering \cite{pal-etal-2022-parameter}, and on the GLUE benchmark \cite{https://doi.org/10.48550/arxiv.2108.02340,ruckle-etal-2021-adapterdrop},
In spite of all its advantages and a large research footprint in NLP, parameter-efficient methods remain under-explored in IR.

A recent comprehensive study~\cite{Ding2022DeltaTA} categorises parameter efficient transfer learning into 3 categories:  1) Addition based 2) Specification based 3) Reparameterization based. Addition based methods insert intermediate modules into the pretrained model. The newly added modules are adapted to the downstream task while keeping the rest of the pretrained model frozen. The modules can be added vertically by increasing the model depth as observed in Houlsby Adapters \cite{conf/icml/HoulsbyGJMLGAG19} and Pfeiffer Adapters \cite{Pfeiffer2020MADXAA}. Houlsby Adapters insert small bottle-neck layers after both the multi-head attention and feed-forward layer of the each transformer layer which are optimized for NLP tasks on GLUE benchmark. Pfeiffer Adapter inserts the bottle-neck layer after only the feed-forward layer and has shown comparable effectiveness to fine-tuning on various NLP tasks. Prompt-based adapter methods such as Prefix-tuning \cite{li-liang-2021-prefix} prepend continuous task-specific vectors to the input sequence which are optimized as free-parameters. Compacter \cite{karimi2021parameter-efficient} hypothesizes that the model can be optimized by learning transformations of the bottle-neck layer in a low-rank subspace leading to less parameters. 

Specification based methods fine-tune only a subset of pretrained model parameters to the task-at-hand while keeping the rest of the model frozen. The fine-tuned model parameters can be only the bias terms as observed in BitFit \cite{ben-zaken-etal-2022-bitfit}, or only cross-attention weights as in the case of Seq2Seq models with X-Attention \cite{gheini-etal-2021-cross}. Re-parameterization methods transform the pretrained weights into parameter efficient form during training. This is observed in LoRA \cite{hu2021lora} which optimises rank decomposition matrices of pretrained layer while keeping the original layer frozen.  

Recent studies exploring parameter efficient transfer learning for Information Retrieval show promising results of such techniques for dense retrieval models \cite{10.1145/3485447.3511978,DBLP:journals/corr/abs-2204-02292,DBLP:journals/corr/abs-2208-09847,https://doi.org/10.48550/arxiv.2207.07087}. \cite{10.1145/3485447.3511978} studies parameter efficient prefix-tuning,  \cite{li-liang-2021-prefix} and LoRA \cite{hu2021lora} on bi-encoder and cross-encoder dense models. Additionally, they combine the two methods by sequentially optimizing one method for \emph{m} epochs, freezing it and optimizing the other for \emph{n} epochs. Their studies show that while cross-encoders with LoRA and LoRA+(50\% more parameters compared to LoRA) outperform fine-tuning with TwinBERT \cite{10.1145/3340531.3412747} and ColBERT \cite{10.1145/3397271.3401075}, parameter-efficient methods \textit{do not outperform fine-tuning} for bi-encoders across all datasets. \cite{DBLP:journals/corr/abs-2204-02292} uses parameter-efficient techniques such as  Sparse Fine-Tuning Masks and Adapters for multilingual and cross-lingual retrieval tasks with rerankers. They train language adapters with Masked Language Modeling (MLM hereafter) task and then task-specific retrieval adapters. This enables the fusion of reranking adapter trained with source language data together with the language adapter of the target language. Concurrent to our work, \cite{https://doi.org/10.48550/arxiv.2207.07087} studies  parameter-efficient prompt tuning techniques such as Prefix tuning and P-tuning v2, specification based methods such as BitFit and adapter-tuning with Pfeiffer Adapters on late interaction bi-encoder models such as Dense Passage Retrieval \cite{karpukhin-etal-2020-dense} and ColBERT. They are motivated by cross-domain generalization of dense retrievals and achieve better results with P-tuning compared to fine-tuning on the BEIR benchmark. \cite{DBLP:journals/corr/abs-2208-09847} studies various 
parameter-efficient tuning procedures at both retrieval and re-
ranking stages. They conduct a comprehensive study of parameter-efficient techniques such as BitFit, Prefix-tuning, Adapters, LoRA, MAM adapters with dense bi-encoders and cross-encoders with BERT-base as the backbone model. Their parameter-efficient techniques achieve comparable effectiveness to fine-tuning on top-20 retrieval accuracy and marginal gains on top-100 retrieval accuracy.


Compared to prior works, our experiments first study the use of adapters for state of the art sparse models such as SPLADE, contrary to previous work that studied dense bi-encoder models\footnote{To the best of our knowledge the only work involving SPLADE and adapters/freezing layers is~\cite{yang2021sparsifying}, which found that freezing the embeddings improves effectiveness.}. Furthermore, our results show improvements compared to the previous studies. We also studied the case of using distinct adapters for query and document encoders in a ``bi-adapter'' setting where the same pretrained backbone model is used by both the query and the document encoder but different adapters are trained for the queries and documents. Secondly, we address another research questions ignored by previous work, which is efficient domain adaptation\footnote{Here we use adaptation as further finetuning on the target domain.} for neural first stage rankers. We start from a trained neural ranker and study adaptation  with adapters on a different domain, such as the ones present in the BEIR benchmark. Finally, we also study parameters sharing between rerankers and first stage rankers using adapters, which to our knowledge has not been studied yet.


\section{Parameter-Efficient Retrieval with Adapters}
\label{sec:previous_work}
In this section, we first present the self-attention used in transformers and how the adapters we use for our experiments interact with them. We then introduce the models used for first stage ranking and reranking.

\subsection{Self-Attention Transformer Layers}

Large pretrained language models are based on the transformer architecture composed of $N$ stacked transformer layers . Each transformer layer comprises of a fully connected feed-forward module and a multi-headed self attention module. Each attention layer has a function of query matrix $(Q \in R^{n X d_k})$, a key matrix and a value matrix. The attention can be formally written as:
\begin{equation}
    A(Q, K, V) = softmax(\frac{QK^T}{\sqrt{d_k}})V
\end{equation}
where the query $Q$, key $K$ and value $V$ are parameterized by  weight matrices $W_q \in R^{nXd_k}$,  $W_k \in R^{nXd_k}$, and  $W_v \in R^{nXd_v}$, as $Q = XW_q$, $K = XW_k$ and $V = XW_v$. Each of the $N$ heads has its respective $Q_i$, $V_i$ and $K_i$ weights and its corresponding attention $A_i$. The feed-forward layer takes as input a transformation of the concatenation of the $N$ attentions as:
\begin{equation}
    FFN(x) = \sigma(XW_1 + b_1)W_2 + b_2
\end{equation}
where $\sigma(.)$ is the activation function. A residual connection is further added after each attention layer and feed-forward layer.

\subsection{Adapters}
\begin{figure}[th!]
\centering
    \includegraphics[width=0.25\linewidth]{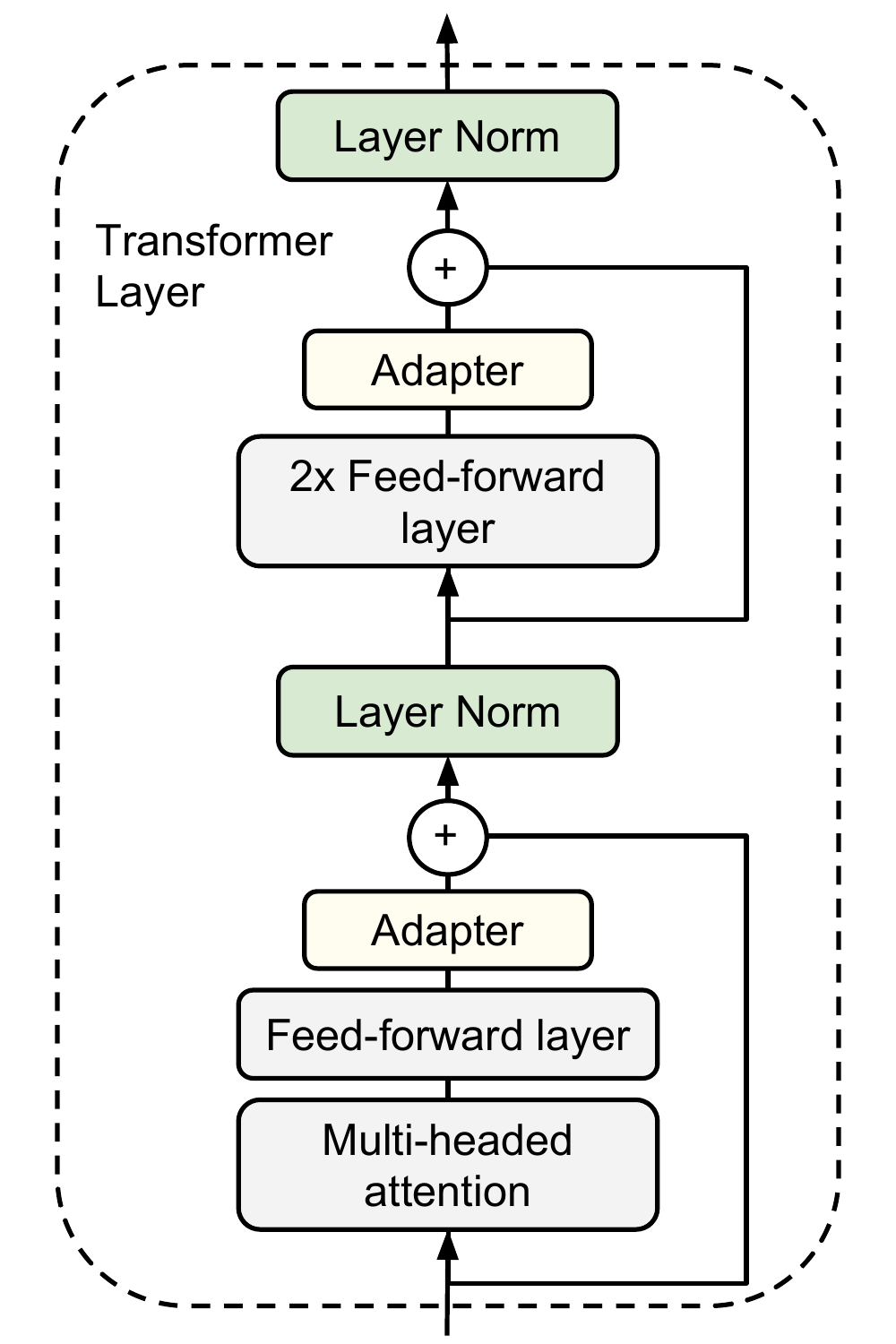}
    \includegraphics[width=.25\linewidth]{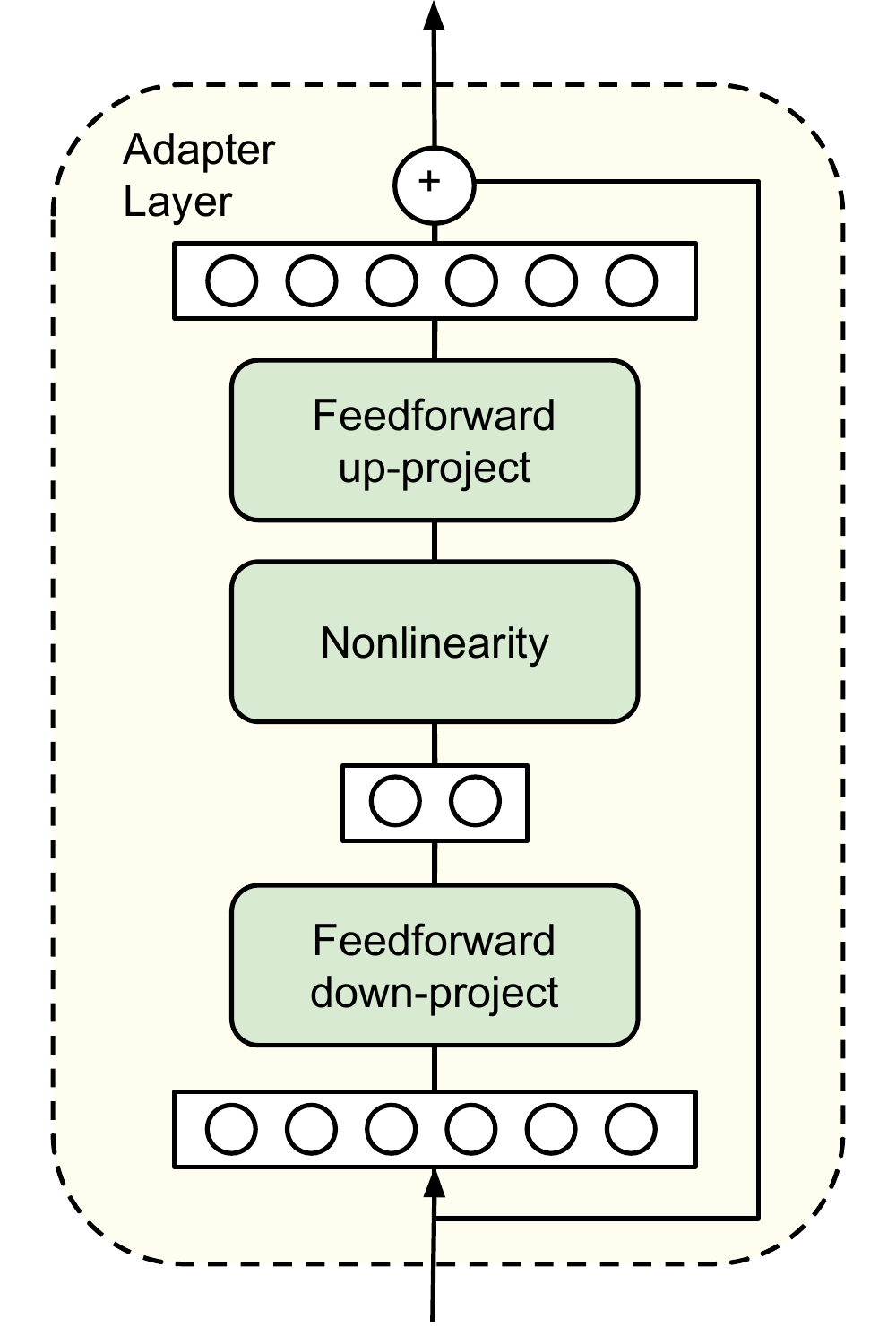}
\caption{Houlsby Adapter, image from the original paper~\cite{conf/icml/HoulsbyGJMLGAG19}}
\label{fig:architecture}
\end{figure}

In this paper, we focus on the Houlsby adapter~\cite{conf/icml/HoulsbyGJMLGAG19}, which as described in Section~\ref{sec:previous_work} can be considered an additive adapter and is depicted in Figure~\ref{fig:architecture}. An additive adapter inserts trainable parameters in addition to the aforementioned transformer layers. The added modules form a bottle-neck architecture with a down-projection, an up-projection and a non-linear transformation. The size of the bottle-neck controls the number of training parameters in an adapter layer. Additionally, a residual connection is applied across each adapter layers. Finally, a layer normalization is added after each transformer sublayer. Formally, this is defined as:
\begin{equation}
    x = f(hW_{down})W_{up} + x
\end{equation}
where $x \in R^d$ is the input to the adapter layer, $W_{down} \in R^{dXr}$ is the down projection matrix transforming input $x$ into bottle-neck dimension $d$, $W_{up} \in R^{rXd}$ is the up projection matrix transforming the bottle-neck representation back to the d-dimensional space. Each adapter layer is initialized with a near-identity weights to enable stable training.

\subsection{Neural Sparse First Stage Retrievers}
Neural sparse first stage retrievers learn contextualized representations of documents and queries in a sparse high-dimensional latent space. In this work, we focus on SPLADE sparse retriever~\cite{10.1145/3477495.3531857,lassance2022efficiency}, which uses both $L_1$ and \emph{FLOPS} regularizations to force sparsity. We freeze the pretrained language model while training the adapter layers. SPLADE predicts term weights of each vocabulary token $j$ with respect to an input token $i$ as:
\begin{equation}
   w_{ij} = transform(h_i)^TE_j + b_j \qquad j \in {1, ..., |V|}
\end{equation}
where $E_j$ is the $j^{th}$ vocabulary token embedding, $b_j$ is it's bias, $h_i$ is $i^{th}$ input token embedding, $transform(.)$ is a linear transformation followed by GeLU activation and LayerNorm. The final term importance for each vocabulary term $j$ is obtained by taking the maximum predicted weights over the entire input sequence of length $n$, after applying a log-saturation effect:
\begin{equation}
w_j = \max_{n}\; \log (1 + ReLU(w_{ij}))
\label{eq:token_rep}
\end{equation}
Given a query $q_i$, the ranking score $s$ of a document $d$ is defined by the degree to which it is relevant to $q$ obtained as a dot product $s(q, d) = w(q). w(d)$. The learning objective is to discriminate representations obtained from Equation \ref{eq:token_rep} of a relevant document $d^+$ and non-relevant hard-negatives $d^-$ obtained from BM25 and in-batch negatives $d^-_{i,j}$ by minimizing the contrastive loss: 
\begin{equation}
\label{eq:contrastive}
    L = -log \frac{e^{s(q_i, d^+)}}{e^{s(q_i, d^+_i)} + e^{s(q_i, d^-_i)} + \sum_j e^{s(q_i, d^-_{i,j})}} 
\end{equation}

SPLADE can be further improved with distillation. The learning objective here is to minimize the MarginMSE~\cite{10.1145/3477495.3531857} loss: mean-squared-error between the positive negative margins of a cross-encoder teacher and the student:
\begin{equation}
    L = MSE(M_s(q_i, d^+) - M_s(q_i, d^-),M_t(q_i, d^+) - M_t(q_i, d^-))
    \label{eq:marginMSE}
\end{equation}
where $MSE$ is mean-squared error, $M_t$ is the teacher's margin and $M_s$ is the student's margin. The final objective optimizes either of the objective in Equation \ref{eq:contrastive} or \ref{eq:marginMSE}  with regularization losses:
\begin{align} 
    \Lb_{SPLADE} &= L + \lambda_q\; \Lb_1 + \lambda_d\; \Lb_{FLOPS} \\
 where\   \Lb_{FLOPS} &= \sum_{j\in V}{\widehat{a^2_j}} = \sum_{j \in V} (\frac{1}{N} \sum_{i=1}^N w_j^{d_i})
\end{align}

The Flops regularizer is a smooth relaxation of the average number of floating-point operations necessary to compute the score of a document, and hence directly related to the retrieval time. It is defined using as a continuous relaxation of the activation (i.e. the term has a non zero weight) probability $a_j$ for token $j$, and estimated for documents $d$ in a batch of size $N$
by $\widehat{a^2_j}$.

\textbf{Retrieval Flops:} SPLADE also reports the retrieval flops (noted R-FLOPS), i.e., the number of floating point operations on the inverted index to return the list of documents for a given query.
The R-FLOPS metric is defined by an estimation of the average number of floating-point operations between a query and a document which is defined as the expectation $\mathbb{E}_{q,d} \left[ \sum_{j \in V} p_j^{(q)}p_j^{(d)} \right]$ where $p_j$ is the activation probability for token $j$ in a document $d$ or a query $q$. It is empirically estimated from a set of approximately $100$k development queries, on the MS MARCO collection. It is thus an indication of the inverted index sparsity and of the computational cost for a sparse model (which is different from the inference ie forward cost of the model)

\subsection{Cross-Encoding Rerankers}

Another way to use PLMs for neural retrieval is to use what is called ``cross-encoding''~\cite{yates2021pretrained}. In this case, both query and document are concatenated before being provided to the network and the score is directly computed by the network. The cross-encoding procedure allows for networks that are much more effective, but this effectiveness comes with a cost on efficiency as the retrieval procedure now has to go through the entire network for each query document pair, instead of being able to precompute document representations and only go through the network for the query representation. The models are trained with a contrastive loss as seen in equation~\eqref{eq:contrastive} that aims to maximize the score of the true query/document pair compared to a BM25 negative query/document pair, without using in-batch negatives.

\section{Experimental Setting and Results}
We use the SPLADE github repository\footnote{https://github.com/naver/splade} to implement our modifications and followed the standard procedure to train SPLADE models. We implement our SPLADE models using an $L_1$ regularization for the query, and $FLOPS$ regularization for the document following ~\cite{lassance2022efficiency}. Unless otherwise stated, the document regularization weight $\lambda_d$ is set to $9\mathrm{e}{-5}$ and the query regularization weight $\lambda_q$ to $5\mathrm{e}{-4}$ to train all variants of Adapters-SPLADE. In order to mitigate the contribution of the regularizer at the early stages of training, we follow \cite{paria2019minimizing} and use a scheduler for $\lambda$, quadratically increasing $\lambda$ at each training iteration, until the $50k$ step. We use a learning rate of $8\mathrm{e}{-5}$, a batch size of 128, a linear scheduler and warmup step of 6000. We set the maximum sequence length to $256$. We train for 300k iterations and keep the best checkpoint using MRR@10 on the validation set. We use a bottle-neck reduction factor of $16$ (i.e. $16$ times smaller) for all adapter layers. We use PyTorch \cite{paszke2019pytorch}, Huggingface Transformers \cite{wolf2020transformers} and AdapterHub \cite{Beck2022AdapterHubPS} to train all models on 4 Tesla V100 GPUs with 32GB memory. We compute statistical significance with $p\le 0.05$ using the Student's t-test and use superscripts to identify statistical significance for almost all measures safe for metrics related to BEIR\footnote{due to lack of standard procedure.}.

\subsection{RQ1: Adapters-SPLADE}
\label{sec:splade_monoencoders}
We study 2 different settings of encoding with adapters. The first called \texttt{adapter}, is a mono-encoder setup where the query and document shars a single encoder. The adapter layers are optimized with both the input sequences keeping the PLM frozen. The second setting inspired by the work on \cite{lassance2022efficiency}, is a bi-encoder setup which separates query and document encoders by training distinct query and document adapters on a shared frozen PLM. We call this setting \texttt{bi-adapter}. This setting not only benefits from optimizing exclusive adapters for input sequence type (different lengths of query/document, etc.), it is also possible to use smaller PLMs for the queries instead of sharing PLM weights. We explore different backbone PLMs: \texttt{DistilBERT} and \texttt{CC+MLM Flops}, a pretrained PLM of cocondenser trained on the masked language model (MLM) task using the FLOPS regularization in order to make it easier to work with SPLADE, introduced in~\cite{lassance2022efficiency}. We trained and evaluated Adapter-SPLADE models on the MS MARCO passage ranking dataset \cite{nguyen2016ms} in full ranking setting. The results for finetuning with BM25 triplets are available in Table \ref{tab:spalde_adapters_results}, whereas in Table~\ref{tab:splade_adapters_results_distill} we make available the results of training models with distillation. For distillation, we use hard-negatives and scores generated by a cross-encoder reranker\footnote{\url{https://huggingface.co/cross-encoder/ms-marco-MiniLM-L-6-v2}} and the MarginMSE loss as described in \cite{10.1145/3477495.3531857} and set $\lambda_d$ to $1\mathrm{e}{-2}$ and $\lambda_q$ to $9\mathrm{e}{-2}$. 

\begin{table}
\caption{Finetuning and adapter-tuning comparison using BM25 triplets for training.}
\centering
\resizebox{\textwidth}{!}{
\begin{tabular}{c|c|c|cc|c|c|c|c}
\toprule
\multirow{2}{*}{\textbf{Model}} & \multirow{2}{*}{\textbf{\#}} &  \multirow{2}{*}{\textbf{Method}} & \multicolumn{2}{c|}{\textbf{MS MARCO dev}} & \multicolumn{1}{c|}{\textbf{TREC DL 2019}} & \multicolumn{1}{c|}{\textbf{TREC DL 2020}} & \multirow{2}{*}{\textbf{R-Flops}} & \multirow{2}{*}{\textbf{\makecell{Training \\ Params}}} \\
& & &  MRR@10 & R@1000 & NDCG@10 & NDCG@10 & & \\
\midrule

\multirow{3}{*}{DistilBERT} & a &finetuning &0.346\hphantom{$^{bcdef}$} &0.963\hphantom{$^{bcdef}$} &0.692\hphantom{$^{bcdef}$} &0.677\hphantom{$^{bcdef}$} & 1.43 & 100\%\\
& b &adapter &0.351\hphantom{$^{acdef}$} &0.968$^{a}$\hphantom{$^{cdef}$} &0.711\hphantom{$^{acdef}$} &0.676\hphantom{$^{acdef}$} & 1.44 & 2.23\%\\
& c &bi-adapter &0.352\hphantom{$^{abdef}$} &0.967$^{a}$\hphantom{$^{bdef}$} &0.690\hphantom{$^{abdef}$} &0.666\hphantom{$^{abdef}$} & 0.74 & 2.23\% \\ \midrule
\multirow{3}{*}{\scriptsize{\makecell{CC + \\MLM FLOPS}}} & d &finetuning &0.366$^{abc}$\hphantom{$^{ef}$} &0.977$^{abc}$\hphantom{$^{ef}$} &\textbf{0.712}\hphantom{$^{abcef}$} &0.684\hphantom{$^{abcef}$} & 1.09 & 100\% \\
& e &adapter &\textbf{0.376}$^{abcd}$\hphantom{$^{f}$} &\textbf{0.980}$^{abcdf}$\hphantom{} &0.712\hphantom{$^{abcdf}$} &0.688\hphantom{$^{abcdf}$} & 0.8 & 2.23\%\\
& f &bi-adapter &0.372$^{abc}$\hphantom{$^{de}$} &0.976$^{abc}$\hphantom{$^{de}$} &0.701\hphantom{$^{abcde}$} &\textbf{0.700}\hphantom{$^{abcde}$} & 0.37 & 2.23\% \\

\bottomrule
\end{tabular}
}
\label{tab:spalde_adapters_results}
\end{table}

\begin{table}
\caption{Finetuning and adapter-tuning comparison using distillation training.}
\centering
\resizebox{\textwidth}{!}{
\begin{tabular}{c|c|c|cc|c|c|c|c}
\toprule
\multirow{2}{*}{\textbf{Model}} & \multirow{2}{*}{\textbf{\#}} &  \multirow{2}{*}{\textbf{Method}} & \multicolumn{2}{c|}{\textbf{MS MARCO dev}} & \multicolumn{1}{c|}{\textbf{TREC DL 2019}} & \multicolumn{1}{c|}{\textbf{TREC DL 2020}} & \multirow{2}{*}{\textbf{R-Flops}} & \multirow{2}{*}{\textbf{\makecell{Training \\ Params}}} \\
& & &  MRR@10 & R@1000 & NDCG@10 & NDCG@10 & & \\
\midrule

\multirow{2}{*}{DistilBERT} & a &finetuning &0.371\hphantom{$^{bcd}$} &0.979$^{b}$\hphantom{$^{cd}$} &0.727\hphantom{$^{bcd}$} &0.711\hphantom{$^{bcd}$} & 3.93 & 100\% \\
& b &adapter &0.373\hphantom{$^{acd}$} &0.975\hphantom{$^{acd}$} &0.728\hphantom{$^{acd}$} &0.716\hphantom{$^{acd}$} & 1.86 & 2.16\% \\ \midrule
\multirow{2}{*}{\scriptsize{\makecell{CC + \\MLM FLOPS}}} & c &finetuning &0.388$^{ab}$\hphantom{$^{d}$} &0.982$^{ab}$\hphantom{$^{d}$} &0.734\hphantom{$^{abd}$} &\textbf{0.732}\hphantom{$^{abd}$} & 4.38 & 100\% \\
& d &adapter &\textbf{0.390}$^{ab}$\hphantom{$^{c}$} &\textbf{0.983}$^{ab}$\hphantom{$^{c}$} &\textbf{0.740}\hphantom{$^{abc}$} &0.729\hphantom{$^{abc}$} & 2.34 & 2.16\% \\

\bottomrule
\end{tabular}
}
\label{tab:splade_adapters_results_distill}
\end{table}

To study efficiency-effectiveness trade-off of Adapters-SPLADE, we compare effectiveness, R-FLOPS size and number of training parameters of adapter-tuned models with their baseline finetuned counterparts having the same backbone PLM. \cite{paria2019minimizing} first showed that R-FLOPs reduction is a reasonable measure of retrieval speed. R-FLOPS measure the average number of floating-point operations needed to compute a document score during retrieval. A sparse embedding and subsequently lower FLOP achieves a retrieval speedup of the order of $1/p^2$ over an inverted index where $p$ is the probability of each document embedding dimension being non-zero. 

Overall, we observe, from Table \ref{tab:spalde_adapters_results} and \ref{tab:splade_adapters_results_distill}, all variants of adapter-tuned SPLADE outperform all baseline fine-tuned counterparts on MS MARCO and TREC DL 2019. The distilled cocondenser with MLM mono-encoder model is the highest performing with an MRR@10 score of $0.390$ and R@100 of $0.983$. The difference in effectiveness between the mono-encoder and bi-encoder adapter-tuning is marginal and depends on the PLM. Most noteworthy, we also observe that the R-FLOPS are lower for adapter-tuned models indicating sparser representation than the fine-tuned counterparts. This is more pronounced in the adapter-tuned models with distillation. Finally, the \textbf{bi-adapter} models have even lower R-FLOPS than the mono-encoder settings, which shows that for the same effectiveness the bi-adapters models are more efficient and sparse. We also observe that the number of training parameters is only $2.23\%$ of the total model parameters for \emph{triplets} training (1.5M/67M for mono-adapter \texttt{DistilBERT}, 3M/135M for bi-adapter \texttt{DistilBERT}, 2M/111M for \texttt{CC + MLM FLOPS}) and $2.16\%$ for the distillation process (1.5M/67M for mono-adapter \texttt{DistilBERT}, 2M/111M for \texttt{CC + MLM FLOPS}).  This has direct consequence in low-hardware setting where adapters with lower number of number of training parameters and gradients can be trained on a smaller GPU(such as 24GB P40) but full finetuning is infeasible. Overall, there is a clear advantage in using Adapter-SPLADE over finetuning, which differs from the previous results on dense adapters \cite{10.1145/3485447.3511978}.

We also evaluate with the full BEIR benchmark [41] comprising of 18 different datasets to measure generalizability of IR models with zero-shot effectiveness on out-of-domain data. The results are listed in Table \ref{tab:max_beir}. We observe from that in the mono-adapter Triplets training, \texttt{adapter} outperforms finetuning on mean nDCG@10 with the highest gap in arguana. With \texttt{CC+MLM Flops} as the backbone model, finetuning and adapter-tuning performs similarly. However,  adapter scores drop on models trained with distillation. This can be attributed to the adapter representations being sparser compared to the finetuned models. As depicted by the R-FLOPS in Table \ref{tab:spalde_adapters_results}, adapter-tuned \texttt{DistilBERT} has less than half the number of R-FLOPS than its finetuned counterpart whereas \texttt{CC+MLM Flops} finetuned model has approximately $1.87$ times the number of R-FLOPS of the adapter-tuned model. This reflects in model representation capacity in 0-shot setting in Table \ref{tab:splade_beir}. However, as discussed in Section \ref{sec:domain_adaptation}, adapters are well suited for domain adaptation when trained on out-of-domain datasets keeping the backbone retriever intact and free from catastrophic forgetting.

\begin{table}[ht!]
    \centering
    \caption{nDCG@10 score comparison on the BEIR zero-shot evaluation}
    \label{tab:max_beir}
\resizebox{\textwidth}{!}{
\begin{tabular}{l|cc|cc|cc|cc}
\toprule
\multirow{3}{*}{Datasets} & \multicolumn{4}{c|}{Triplets training} & \multicolumn{4}{c}{Distillation training} \\
&  \multicolumn{2}{c|}{DistilBERT} & \multicolumn{2}{c|}{CC + MLM FLOPS} &  \multicolumn{2}{c|}{DistilBERT} & \multicolumn{2}{c}{CC + MLM FLOPS}  \\
 &  finetuning &  adapter &  finetuning & adapter & finetuning &  adapter &  finetuning & adapter \\
\midrule
arguana          &                                    0.298 &                                  0.364 &                                  0.427 &                                0.388 &                                           0.513 &                                         0.443 &                                         0.463 &                                       0.433 \\
climate-fever    &                                    0.167 &                                  0.172 &                                  0.180 &                                0.187 &                                           0.202 &                                         0.197 &                                         0.229 &                                       0.202 \\
dbpedia-entity   &                                    0.379 &                                  0.392 &                                  0.388 &                                0.401 &                                           0.419 &                                         0.417 &                                         0.438 &                                       0.432 \\
fever            &                                    0.730 &                                  0.734 &                                  0.724 &                                0.722 &                                           0.773 &                                         0.757 &                                         0.792 &                                       0.773 \\
fiqa             &                                    0.295 &                                  0.289 &                                  0.317 &                                0.320 &                                           0.332 &                                         0.314 &                                         0.342 &                                       0.337 \\
hotpotqa         &                                    0.626 &                                  0.647 &                                  0.650 &                                0.603 &                                           0.687 &                                         0.670 &                                         0.687 &                                       0.629 \\
nfcorpus         &                                    0.318 &                                  0.321 &                                  0.331 &                                0.333 &                                           0.335 &                                         0.335 &                                         0.340 &                                       0.344 \\
nq               &                                    0.481 &                                  0.482 &                                  0.506 &                                0.523 &                                           0.522 &                                         0.508 &                                         0.539 &                                       0.544 \\
quora            &                                    0.819 &                                  0.810 &                                  0.821 &                                0.806 &                                           0.825 &                                         0.722 &                                         0.841 &                                       0.552 \\
scidocs          &                                    0.143 &                                  0.150 &                                  0.151 &                                0.153 &                                           0.154 &                                         0.147 &                                         0.152 &                                       0.153 \\
scifact          &                                    0.614 &                                  0.611 &                                  0.658 &                                0.669 &                                           0.687 &                                         0.658 &                                         0.690 &                                       0.673 \\
trec-covid       &                                    0.694 &                                  0.684 &                                  0.668 &                                0.689 &                                           0.703 &                                         0.728 &                                         0.700 &                                       0.713 \\
webis-touche2020 &                                    0.270 &                                  0.255 &                                  0.277 &                                0.274 &                                           0.260 &                                         0.258 &                                         0.294 &                                       0.290 \\ \midrule
mean             &                                   0.449 &                                 0.455 &                                 0.469 &                               0.467 &                                          0.493 &                                        0.473 &                                        0.500 &                                      0.467 \\
\bottomrule
\end{tabular}
}
\label{tab:splade_beir}
\end{table}

\subsection{RQ2: Adapter Layer Ablation}
Furthermore, we perform extensive adapter layer ablation by progressively removing adapter layers from the early layers of the encoder. Doing so results in $n$ separate models for each layer ablation setting. The frozen pretrained model for our ablation studies is \texttt{DistilBERT} in a mono-encoder setting where the same instance of the encoder is used to encode both the document and the query, which is the same configuration as the adapter method in Table~\ref{tab:spalde_adapters_results}.  This results in a total of $6$ configurations for the ablation study corresponding to the 6 adapter layers after each pretrained transformer layer. The final experimental setting removes all $6$ adapter layers ($0-5$) and fine-tunes only the language model head. 
 \begin{table}[t!]
\caption{Adapter layer Ablation with \texttt{adapters} on \texttt{DistilBERT} PLM.}
\centering
\resizebox{\textwidth}{!}{
\begin{tabular}{c|c|cc|c|c|c|c|c|c}
\toprule
\multirow{2}{*}{\textbf{\#}} & \multirow{2}{*}{\textbf{\makecell{Adapters\\ Removed}}} & \multicolumn{2}{c|}{\textbf{MS MARCO dev}} & \textbf{TREC DL 2019} & \textbf{TREC DL 2020} & 
\textbf{BEIR} & 
\multirow{2}{*}{\textbf{R-Flops}} & \multirow{2}{*}{\textbf{\makecell{Training \\ Params}}} & \multirow{2}{*}{\textbf{\makecell{Training\\Time(Hrs)}}} \\
& & MRR@10 & R@1000 & NDCG@10 & NDCG@10  & NDCG@10  & & \\
\midrule
a &None &\textbf{0.351}$^{cdefg}$\hphantom{$^{b}$} &\textbf{0.968}$^{defg}$\hphantom{$^{bc}$} &\textbf{0.711}$^{fg}$\hphantom{$^{bcde}$} &0.676$^{g}$\hphantom{$^{bcdef}$} & 0.455 & 1.44 & 2.23\% & 34.42\\ \midrule
b &0 &0.348$^{defg}$\hphantom{$^{ac}$} &0.967$^{efg}$\hphantom{$^{acd}$} &0.708$^{fg}$\hphantom{$^{acde}$} &0.674$^{g}$\hphantom{$^{acdef}$} & 0.458 & 1.27 & 2.01\% & 32.23 \\
c &0-1 &0.344$^{efg}$\hphantom{$^{abd}$} &0.968$^{efg}$\hphantom{$^{abd}$} &0.709$^{fg}$\hphantom{$^{abde}$} &\textbf{0.699}$^{abdefg}$\hphantom{} & 0.459 & 1.34 & 1.80\%  & 28.55 \\
d &0-2 &0.341$^{efg}$\hphantom{$^{abc}$} &0.966$^{efg}$\hphantom{$^{abc}$} &0.703$^{fg}$\hphantom{$^{abce}$} &0.665$^{g}$\hphantom{$^{abcef}$} & 0.459 & 1.36 & 1.59\% & 26.70 \\
e &0-3 &0.325$^{fg}$\hphantom{$^{abcd}$} &0.962$^{fg}$\hphantom{$^{abcd}$} &0.689\hphantom{$^{abcdfg}$} &0.660$^{g}$\hphantom{$^{abcdf}$} & 0.455 & 1.50 & 1.37\% & 24.18\\
f &0-4 &0.318$^{g}$\hphantom{$^{abcde}$} &0.956\hphantom{$^{abcdeg}$} &0.659\hphantom{$^{abcdeg}$} &0.663$^{g}$\hphantom{$^{abcde}$} & 0.455 & 1.27 & 1.15\% & 22.51\\
g &0-5 &0.312\hphantom{$^{abcdef}$} &0.955\hphantom{$^{abcdef}$} &0.660\hphantom{$^{abcdef}$} &0.617\hphantom{$^{abcdef}$} & 0.449 & 2.78 & 0.90\% & 21.35\\
\bottomrule
\end{tabular}
}
\label{tab:ablation_msmarco}
\end{table}

 We note that such an experiment (dropping adapter layers from transformer models) has been studied in NLP \cite{ruckle-etal-2021-adapterdrop} and was shown to improve both training and inference time while retaining comparable effectiveness. We report the effectiveness of each adapter ablation setting on MS MARCO, TREC DL 2019 and TREC DL 2020 in Table \ref{tab:ablation_msmarco}. We actually observe gradual performance drop for MS MARCO and TREC DL datasets as the training parameters decrease with the progressive removal of adapter layers as shown in Table \ref{tab:ablation_msmarco}. The drop is significantly higher (a drop of $0.25$ MRR score) when layers are removed from the second half of the model ( $\ge 0-3$). This phenomenon is consistent with studies in NLP \cite{pal-etal-2022-parameter,ruckle-etal-2021-adapterdrop} that task-specific information is stored in the later layers of the adapters. For the BEIR datasets, this effectiveness drop is not as evident until all adapters but the language model head is removed (configuration $0-5$). The last configuration also has less sparsity as observed from the R-FLOPS size of $2.78$ compared to the other configurations. We also observe that the training time drops proportional to the drop in adapter layers. The training time for adapter-tune without any drop in adapter layers is $34.42$ hours on 4 Tesla V100 GPUS for $150,000$ iterations, and it drops to $26.70$ hours with only $1\%$ drop in MRR with the first $0-2$ adapter layers dropped. The lowest training time is $21.35$ hours with a drop of $3.2\%$ in MRR for the configuration with all adapters dropped but the language model head.
 
\subsection{RQ3: Out-of-Domain Dataset Adaptation}
\label{sec:domain_adaptation}
For the next research question, we want to check how adapters compare to full finetuning when adapting a model trained on MSMARCO on a smaller out-of-domain dataset. We evaluate this question under two scenarios: i) BEIR and ii) TripClick. 

\paragraph{\textbf{BEIR}:} On the beir benchmark we use 3 datasets (FEVER, FiQA and NFCorpus) that have training, development and test sets and aim for very different domains and tasks (fact checking , financial QA and bio-medical IR). We start from a pre-finetuned SPLADE model called ``splade-cocondenser-ensembledistil'' made available in~\cite{10.1145/3477495.3531857}. We verify the effectiveness of the models in zero shot and get a first set of hard negatives. These hard negatives are then used to train either via finetuning of all parameters or via the introduction of adapters. The networks are trained for either 10 (FEVER) or 100 epochs (FiQA and NFCorpus), and at the end of each epoch we compute the development set effectiveness. We use the models with the best development set to compute the 1st round test set effectiveness and generate hard negatives that are used for another round of training that we call 2nd round (which repeats the 1st round, starting from the best network of the 1st round and using negatives from the 1st round). 

Results are available in Table~\ref{tab:beir_adapt}. While finetuning is not always able to improve the results over the zero-shot, mostly due to overfitting on the training/dev sets. For example, on fever fine-tuning first makes all representations as it can easily overfit to the training even without using many words and only on the second round of training started using more dimensions. On the other hand, adapter tuning is able to consistently improve the effectiveness over the zero shot and first rounds (even if it does not always perform the best, as is the case on NFCorpus). Overall, we conclude that adapters are more stable than finetuning when finetuning on these specific domains.

\begin{table}
\caption{Domain adaptation comparison on BEIR Datasets}
\label{tab:beir_adapt}
\centering
\resizebox{\textwidth}{!}{
\begin{tabular}{c|c|cc|cc|cc}
\hline
\multirow{2}{*}{\textbf{Dataset}} & \multirow{2}{*}{\textbf{Training}} & \multicolumn{2}{c|}{\textbf{Zero Shot}} &  \multicolumn{2}{c|}{\textbf{1st Round}} & \multicolumn{2}{c}{\textbf{2nd Round}} \\
& &  NDCG@10 & Recall@100 & NDCG@10 & Recall@100 &  NDCG@10 & Recall@100 \\
\hline
\multirow{2}{*}{Fever} & finetuning & \multirow{2}{*}{0.793} & \multirow{2}{*}{0.954} & 0.692 & 0.866 & 0.851  & 0.959 \\
& adapter & & & 0.841 & 0.960  & \textbf{0.881} & \textbf{0.964} \\
\hline
 \multirow{2}{*}{FiQA} & finetuning & \multirow{2}{*}{0.348} & \multirow{2}{*}{0.632} & 0.371 & 0.678 & 0.356 & 0.694 \\
 & adapter  & & & 0.373  & 0.675 & \textbf{0.393} & \textbf{0.711} \\
\hline
\multirow{2}{*}{NFCorpus} & finetuning & \multirow{2}{*}{0.348} & \multirow{2}{*}{0.285} & 0.384 & 0.466  & \textbf{0.403}  & \textbf{0.484} \\
& adapter & & & 0.362 & 0.435 & 0.371  & 0.428  \\
 \hline
\end{tabular}
}
\label{tab:tablesum_results}
\end{table}

\paragraph{\textbf{TripClick}:} Given that in the BEIR benchmark the adapters underperformed finetuning on bio-medical data, we decided to further experiment on a larger bio-medical dataset called TripClick. The TripClick collection \cite{rekabsaz2021fairnessir} contains approximately 1.5 millions MEDLINE documents (title and abstract), and 692,000 queries. The test set is divided into three categories of queries: Head, Torso and Tail (according to their decreasing frequency), which contain 1,175 queries each. For the Head queries, a DCTR click model was employed to created relevance signals, otherwise raw clicks were used. We use the triplets released by \cite{hofstaetter2022tripclick}.
Similarly to the BEIR experiments, we start from the ``splade-cocondenser-ensembledistil'' SPLADE model and fine-tune or adapt-tune it over 100,000 iterations (batch size equal to 100). As shown in Table~\ref{tab:tripclick}, adapter-tuning shows very competitive results, on par with finetuning for head categories (frequent queries), and achieving even better results for the less frequent queries (torso and tail). 

\begin{table}
\caption{Performance of mono-encoder on out-of-domain Tripclick Dataset}
\centering
\resizebox{\textwidth}{!}{
\begin{tabular}{c|c|cc|cc|cc|cc}
\toprule
\multirow{2}{*}{\textbf{\#}} &
\multirow{2}{*}{\textbf{Training}} &  \multicolumn{2}{c|}{\textbf{HEAD (dctr)}}   &\multicolumn{2}{c|}{\textbf{HEAD}} &  \multicolumn{2}{c|}{\textbf{Torso}} & \multicolumn{2}{c}{\textbf{Tail}} \\
         &                          &  NDCG@10 & Recall@100 & NDCG@10 & Recall@100 &  NDCG@10 & Recall@100 &  NDCG@10 & Recall@100  \\
\midrule

a &
Finetuning &
0.218\hphantom{$^{b}$} &
\textbf{0.579}\hphantom{$^{b}$} &
\textbf{0.302}\hphantom{$^{b}$} &
0.523\hphantom{$^{b}$} &
0.219\hphantom{$^{b}$} &
\textbf{0.679}\hphantom{$^{b}$} &
0.238\hphantom{$^{b}$} &
\textbf{0.722}\hphantom{$^{b}$} \\
b &
Adapter &
\textbf{0.219}\hphantom{$^{a}$} &
0.578\hphantom{$^{a}$} &
0.299\hphantom{$^{a}$} &
\textbf{0.526}\hphantom{$^{a}$} &
\textbf{0.229}$^{a}$\hphantom{} &
0.679\hphantom{$^{a}$} &
\textbf{0.253}$^{a}$\hphantom{} &
0.720\hphantom{$^{a}$} \\
\bottomrule

\end{tabular}
}
\label{tab:tripclick}
\end{table}

\subsection{RQ4: Knowledge Sharing between Rerankers and First stage Rankers}
The final research question explores sharing knowledge between rerankers and first-stage rankers. We explore this with transforming first stage rankers into rerankers. First, we tune the pretrained \texttt{DistilBERT} for reranking task as a baseline for both finetuning and adapter-tuning. We then test transforming both sparse (\texttt{splade-cocondenser}) and dense (\texttt{tct\_colbert-v2-msmarco}) first stage rankers into rerankers, using either fine-tuning or adapter-tuning. To be clear, the cross-encoder is initialized with the weights of the aforementioned first stage models, but the reranker classification head on the \texttt{CLS} token is randomly initialized. Also note that we rerank the top-1k returned from ``splade-cocondenser-ensembledistil'' (represented by ``first stage'' on table).

We compare adapter-tuning with finetuning and display the results in Table~\ref{tab:reranker}. We observe that finetuning the baseline model (\texttt{DistilBERT}) is better than adapter-tuning. When using first stage rankers, results are varied. Dense first stage rerankers were able to learn similarly with both adapter and fine-tuning. However, this was not the case for sparse first stage rankers \\ (\texttt{splade-cocondenser-ensembledistil}). We posit that this may come from two different reasons: i) The SPLADE model does not focus on the CLS representations, but on the MLM head representations of all tokens, thus needing more flexibility; ii) The model has been trained multiple times (initial BERT training, then condenser, then cocondenser and finally SPLADE), while not always using the same precision (fp16 or fp32), which under preliminary analysis seems to have made some parts of the model unusable for cross-encoding without full finetuning. 
Overall, there is slight gain in using the first stage model for the reranker. However, there's no increase in effectiveness of using adapters, we actually see worse effectiveness on all settings.

\begin{table}[t!]
\caption{Knowledge Sharing between first stage rankers and rerankers comparison between finetuning and adapter-tuning.}
\centering
\resizebox{\textwidth}{!}{
\begin{tabular}{c|c|c|c|c|c|c}
\toprule
\multirow{2}{*}{\textbf{Base Model}} & \multirow{2}{*}{\textbf{\#}} & \multirow{2}{*}{\textbf{Training}}  & \textbf{MS MARCO dev} & \textbf{TREC DL 2019} & \textbf{TREC DL 2020} \\
& & &  MRR@10 & NDCG@10 & NDCG@10 \\ \midrule
First Stage & a &None &0.383$^{e}$\hphantom{$^{bcdfg}$} &0.732\hphantom{$^{bcdefg}$} &0.721\hphantom{$^{bcdefg}$} \\ \midrule
\multirow{2}{*}{DistilBERT} & b &finetune &0.396$^{ace}$\hphantom{$^{dfg}$}  &\textbf{0.764}$^{e}$\hphantom{$^{acdfg}$} &0.736\hphantom{$^{acdefg}$} \\
& c &adapter &0.388$^{e}$\hphantom{$^{abdfg}$} &0.737\hphantom{$^{abdefg}$} &0.727\hphantom{$^{abdefg}$} \\ \midrule
\multirow{2}{*}{\texttt{SPLADE++}} & d &finetune &\textbf{0.408}$^{abceg}$\hphantom{$^{f}$} &0.753\hphantom{$^{abcefg}$} &\textbf{0.743}\hphantom{$^{abcefg}$} \\
& e &adapter &0.358\hphantom{$^{abcdfg}$} &0.723\hphantom{$^{abcdfg}$} &0.707\hphantom{$^{abcdfg}$} \\ \midrule
\multirow{2}{*}{\texttt{TCT Colbert v2}} & f &finetune &0.404$^{abce}$\hphantom{$^{dg}$} &0.749\hphantom{$^{abcdeg}$} &0.731\hphantom{$^{abcdeg}$} \\
& g &adapter &0.400$^{ace}$\hphantom{$^{bdf}$} &0.740\hphantom{$^{abcdef}$} &0.739\hphantom{$^{abcdef}$} \\
\bottomrule
\end{tabular}
}
\label{tab:reranker}
\end{table}

\section{Conclusion}
Retrieval models, based on PLM, require finetuning millions of parameters   which makes them memory inefficient and non-scalable for out-of-domain adaptation. This motivates the need for efficient methods to adapt them to information retrieval tasks. In this paper, we examine adapters for sparse retrieval models. We show that with approximately 2\% of training parameters, adapters can be successfully employed for SPLADE models with comparable or even better effectiveness on benchmark IR datasets such as MS MARCO and TREC. We further analyze adapter layer ablation and see a further reduction in training parameters to $1.8\%$ retains effectiveness of full finetuning. For domain adaptation, adapters are more stable and outperform finetuning, which is prone to over-fitting, 
On Tripclick dataset, adapters outperform on precision metrics Torso and Tail queries and performs comparably on Head queries. We explore knowledge transfer between first stage rankers and rerankers as a final study. Adapters underperform  full finetuning when trying to reuse  sparse model to rerankers. Dense first stage rankers perform similarly for adapters and finetuning while sparse first stage rankers is less effective compared to finetuning. We leave this as future work.  As memory-efficient adapters are effective for Splade, we leave for future studying larger sparse models and their generalizability. Finally, an interesting scenario could also be to tackle unsupervised domain adaptation with adapters. 




%
%
\bibliographystyle{splncs04}
%
\bibliography{bibliography}
\end{document}